\documentclass[twoside]{ilcws08}
\usepackage[latin1]{inputenc}
\usepackage[dvips]{graphicx,epsfig,color}
\usepackage{wrapfig,rotating}
\usepackage{amssymb,amsmath,array}
\usepackage{booktabs}
\usepackage{subfigure}

\newcommand{\gsim}{\mbox{ \raisebox{-1.0ex}{$\stackrel{\textstyle >}
{\textstyle \sim}$ }}}
\newcommand{\lsim}{\mbox{ \raisebox{-1.0ex}{$\stackrel{\textstyle <}
{\textstyle \sim}$ }}}

\begin{document}
%

%
%


\setcounter{section}{0}
\setcounter{figure}{0}
\setcounter{table}{0}

\begin{center}
    {\large\bf
ILC phenomenology in a TeV scale radiative seesaw model
 for neutrino mass, dark matter and baryon asymmetry
}\\
\vspace{0.8cm}
        {\bf Mayumi Aoki}$^{(a)}$,
        {\bf Shinya Kanemura}$^{(b)}$,
         {\bf Osamu Seto}$^{(c)}$
         
    \vspace{0.8cm}

    {\it
    $^{(a)}${Institute for Theoretical Physics, Kanazawa University, Kanazawa 920-1192, Japan}\\
    $^{(b)}${Department of Physics, University of Toyama, Toyama 930-8555, Japan}\\
      $^{(c)}${Department of Architecture and Building Engineering, Hokkai-Gakuen University, Sapporo 062-8605, Japan}
 }

    \vspace{0.8cm}

\begin{quote} \small
We discuss phenomenology in a new TeV scale model which would explain neutrino
 oscillation, dark matter, and baryon asymmetry of the Universe
 simultaneously by the dynamics of the extended Higgs sector and TeV-scale
 right-handed neutrinos. 
 Tiny neutrino masses are generated at the three-loop level due to the exact $Z_2$ symmetry,
by which the stability of the dark matter candidate is guaranteed. 
The model provides various discriminative predictions in Higgs 
phenomenology, which can be tested at the Large Hadron Collider
 and the International Linear Collider.
\end{quote}

  \end{center}
  
\section{Introduction}

In spite of the success of the Standard Model (SM) for elementary
particles, it is widely understood that a new model beyond the SM
must be considered to explain the phenomena such as tiny neutrino
masses and their mixing~\cite{lep-data}, the nature of dark
matter (DM)~\cite{wimp} and baryon asymmetry of the
Universe~\cite{sakharov}. 

We here discuss the model in which these problems would be
simultaneously explained by the TeV-scale physics~\cite{aks}.
Tiny neutrino masses are generated at the three-loop level due to an
exact discrete symmetry, by which tree-level Yukawa couplings of neutrinos are prohibited.
The lightest neutral odd state under the discrete symmetry is a
candidate of DM.  
Baryon asymmetry can also be generated at the electroweak phase transition
(EWPT) by additional CP violating phases in the Higgs sector~\cite{ewbg-thdm}.
In this framework, a successful model can be made without contradiction
of the current data.

The original idea of generating tiny neutrino masses via the radiative effect 
has been proposed by Zee~\cite{zee}. 
The extension with a TeV-scale right-handed (RH) neutrino has been discussed in Ref.~\cite{knt},
where neutrino masses are generated at the three-loop level due to the exact $Z_2$
parity, and the $Z_2$-odd RH neutrino is a candidate of DM. This 
has been extended with two RH neutrinos to describe the neutrino data~\cite{kingman-seto}. 
Several models with adding baryogenesis have been considered in Ref.~\cite{ma}.
The following advantages would be in the present model~\cite{aks}:
(a)~all mass  scales are at most at the TeV scale without large hierarchy, 
(b)~physics for generating neutrino masses is connected with that for
DM and baryogenesis, 
(c)~the model parameters are strongly constrained by the current data, so
    that the model provides discriminative predictions which can be tested
    at future experiments.

In the following, we first explain the basic properties of the model,
and discuss its phenomenology, in particular
that at the International Linear Collider (ILC). 
    
\section{Model}

Two scalar isospin doublets with hypercharge $1/2$ ($\Phi_1$ and $\Phi_2$),  
charged singlet fields ($S^\pm$), a real scalar singlet ($\eta$) and two
generation isospin-singlet RH neutrinos ($N_R^\alpha$ with $\alpha=1,
2$) are introduced in our model~\cite{aks}.
We impose an exact $Z_2$ symmetry to generate tiny neutrino masses
at the three-loop level, which we refer as $Z_2$. 
We assign $Z_2$-odd charge to $S^\pm$, $\eta$ and $N_R^\alpha$, while 
ordinary gauge fields, quarks and leptons and Higgs doublets are  $Z_2$ even.
In order to avoid the flavor changing neutral current in a natural way, we impose
another (softly-broken) discrete symmetry ($\tilde{Z}_2$)~\cite{glashow-weinberg}.
We employ so called Type-X Yukawa interaction~\cite{typeX},
where  $\tilde{Z}_2$ charges are assigned such that only $\Phi_1$ couples
to leptons whereas $\Phi_2$ does to quarks~\cite{barger,grossman,typeX2};  
\begin{eqnarray}
 {\cal L}_Y  \!=\! -\!  y_{e_i}^{}  \overline{L}^i \Phi_1 e_R^i
     \! - \! y_{u_i}^{}  \overline{Q}^i \tilde{\Phi}_2 u_R^i
     \! - \! y_{d_i}^{}  \overline{Q}^i \Phi_2 d_R^i + {\rm h.c.}, \label{typex-yukawa}
\end{eqnarray}
where $Q^i$ ($L^i$) is the ordinary $i$-th generation left-handed (LH) quark (lepton)
doublet,  and $u_R^i$ and $d_R^i$ ($e_R^i$) are RH-singlet up- and
down-type quarks (charged leptons), respectively.   
We summarize the particle properties under $Z_2$ and $\tilde{Z}_2$ in Table~\ref{discrete}.

\begin{table}[b]
\begin{center}
\centerline{
  \begin{tabular}{c|ccccc|cc|ccc}
   \hline
   & $Q^i$ & $u_R^{i}$ & $d_R^{i}$ & $L^i$ & $e_R^i$ & $\Phi_1$ & $\Phi_2$ & $S^\pm$ &
    $\eta$ & $N_{R}^{\alpha}$ \\\hline
&&&&&&&&&&\\[-4mm]
$Z_2\frac{}{}$                ({\rm exact}) & $+$ & $+$ & $+$ & $+$ & $+$ & $+$ & $+$ & $-$ & $-$ & $-$ \\ \hline  
&&&&&&&&&&\\[-4mm]
$\tilde{Z}_2\frac{}{}$ ({\rm softly\hspace{1mm}broken})& $+$ & $-$ & $-$ & $+$ &
                       $+$ & $+$ & $-$ & $+$ & $-$ & $+$ \\\hline
   \end{tabular}
 }
  \caption{Particle properties under the discrete symmetries.
 }
  \label{discrete}
\end{center}
\end{table}
The Yukawa coupling in Eq.~(\ref{typex-yukawa}) is different
from that in the minimal supersymmetric SM (MSSM)~\cite{hhg}.
In addition to the usual potential of the two Higgs doublet model (THDM) with
the $\tilde{Z}_2$ parity and that of  the $Z_2$-odd scalars,
we have the interaction terms between $Z_2$-even and -odd scalars:  
\begin{eqnarray}
{\cal L}_{int} = -
 \sum_{a=1}^2 \left(\rho_a |\Phi_a|^2|S|^2 + \sigma_a |\Phi_a|^2
  \frac{\eta^2}{2}\right)
-\sum_{a,b=1}^2\left\{ \kappa \,\,\epsilon_{ab} (\Phi^c_a)^\dagger
                    \Phi_b S^- \eta + {\rm h.c.}\right\},
 \end{eqnarray}
where $\epsilon_{ab}$ is the anti-symmetric tensor with $\epsilon_{12}=1$.
The mass term and the interaction for $N_R^\alpha$ are given by 
\begin{eqnarray}
 {\cal L}_{Y_N^{}} \!= \! \sum_{\alpha=1}^2\!\left\{ \!\frac{1}{2}m_{N_R^\alpha}^{} \overline{{N_R^\alpha}^c} N_R^\alpha
                 -  h_i^\alpha \overline{(e_R^i)^c}
                   N_R^\alpha S^-\! + {\rm h.c.}\!\right\}.
\end{eqnarray} 
Although the CP violating phase in the Lagrangian is
crucial for successful baryogenesis at the EWPT~\cite{ewbg-thdm},
it does not much affect the following discussions. Thus, we neglect it for simplicity.
We later give a comment on the case with the non-zero CP-violating phase. 

As $Z_2$ is exact, the even and odd fields cannot mix.
Mass matrices for the $Z_2$-even scalars are diagonalized as in the
usual THDM by the mixing angles $\alpha$ and $\beta$, where $\alpha$
diagonalizes the CP-even states, and $\tan\beta=\langle \Phi_2^0
\rangle/\langle \Phi_1^0 \rangle$~\cite{hhg}. 
The $Z_2$ even physical states are two CP-even ($h$ and $H$),
a CP-odd ($A$) and charged ($H^\pm$) states.
We here define $h$ and $H$ such that $h$ is always
the SM-like Higgs boson when $\sin(\beta-\alpha)=1$. 

\section{Neutrino Mass, Dark Matter, and Strongly 1st-Order Phase Transition}

\begin{figure}
\begin{center}
 \includegraphics[width=.7\textwidth]{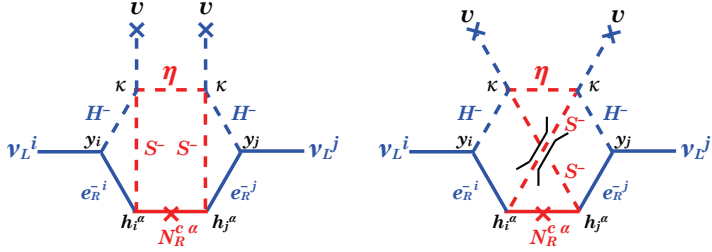}
  \caption{The diagrams for generating tiny neutrino masses. }
  \label{diag-numass}
\end{center}
\end{figure}

The LH neutrino mass matrix $M_{ij}$ is generated by the three-loop diagrams in Fig.~\ref{diag-numass}.
The absence of lower order loop contributions is guaranteed by $Z_2$.
$H^\pm$  and  $e_R^i$ play a crucial role to connect LH neutrinos with the one-loop sub-diagram
by the $Z_2$-odd states.
We obtain
\begin{eqnarray}
M_{ij} = \sum_{\alpha=1}^{2} 
  C_{ij}^\alpha F(m_{H^{\pm}}^{},m_{S^{\pm}}^{},m_{N_R^{\alpha}}^{}, m_\eta), 
\end{eqnarray}
where $C_{ij}^\alpha =
   4 \kappa^2 \tan^2\!\beta 
  (y_{e_i}^{\rm SM} h_i^\alpha) (y_{e_j}^{\rm SM} h_j^\alpha)$ with 
$y_{e_i}^{\rm SM}=\sqrt{2}m_{e_i}/v$ and  $v\simeq 246$ GeV.
  The factor of the three-loop integral function
   $F(m_{H^{\pm}}^{},m_{S^{\pm}}^{},m_{N_R}^{}, m_\eta)$
  includes the suppression factor of $1/(16\pi^2)^3$, whose typical size
  is ${\cal O}(10^{4})$eV.
Magnitudes of $\kappa \tan\beta$ as well as $F$
determine the universal scale of $M_{ij}$, 
whereas variation of $h_i^\alpha$ ($i=e$, $\mu$, $\tau$) 
reproduces the mixing pattern indicated by the neutrino
 data~\cite{lep-data}.
\begin{table}
\begin{center}
  \begin{tabular}{|c||c|c|c|c|c|c|c|}\hline
     Set   & $h_e^1$ & $h_e^2$ & $h_\mu^1$ & $h_\mu^2$ & $h_\tau^1$ & $h_\tau^2$  &
   $B(\mu\!\!\to\!\! e\gamma)$ \\\hline 
  A &  2.0    &  2.0     &  0.041     & -0.020
                   & 0.0012   & -0.0025  & $1.5\!\times \!10^{-12}$  \\\hline 
   B & 2.2     &  2.1     &  0.0087     & 0.037 
                   & -0.0010  &    0.0021  & $7.8\!\times \!10^{-12}$ \\\hline 
   \end{tabular}
\end{center}
 \caption{Values of $h_i^\alpha$ for $m_{H^\pm}^{}
 (m_{S^\pm}^{})=100$ (400) GeV,  
  $m_\eta=50$ GeV, $m_{N_R^1}=m_{N_R^2}=$3.0 TeV for the normal
 hierarchy.  For  Set A (B), 
  $\kappa\tan\beta=29$ (34) and $U_{e3}=0$ (0.14). 
 Predictions on the branching ratio of $\mu\to e
 \gamma$ are also shown.}   \label{h-numass}
 \end{table} 
 
Under the {\it natural} requirement 
$h_e^\alpha \sim {\cal O}(1)$, and taking 
the  $\mu\to e\gamma$ search results into account~\cite{lfv-data},   
we find that $m_{N_R^\alpha}^{} \sim {\cal O}(1)$ TeV, 
$m_{H^\pm}^{} \lsim {\cal O}(100)$ GeV, $\kappa \tan\beta \gsim {\cal
O}(10)$, and $m_{S^\pm}^{}$ being several times 100 GeV. 
On the other hand, the LEP direct search results indicate 
$m_{H^\pm}^{}$ (and $m_{S^\pm}^{}$)  $\gsim 100$ GeV~\cite{lep-data}.  
In addition, with the LEP precision measurement for the $\rho$ parameter,  
possible values uniquely turn out to be  
$m_{H^\pm}^{} \simeq m_{H}^{}$ (or $m_{A}^{}$) $\simeq 100$ GeV
for $\sin(\beta-\alpha) \simeq 1$. 
Thanks to the Yukawa coupling in Eq.~(\ref{typex-yukawa}), such
a light $H^\pm$ is not excluded by the $b \to s \gamma$ data~\cite{bsgamma}.
Since we cannot avoid to include the hierarchy among $y_i^{\rm SM}$,  
we only require $h_i^\alpha y_i \sim {\cal O}(y_e) \sim 10^{-5}$ 
for values of $h_i^\alpha$. 
Our model turns out to prefer the normal hierarchy
scenario. 
Several sets for $h_i^\alpha$ are shown in Table~\ref{h-numass} with the
predictions on the branching ratio of $\mu\to e\gamma$ 
assuming the normal hierarchy\footnote{The predictions for $\mu\to
e\gamma$ shown here are corrected ones from those in Ref.~\cite{aks}.}
.
%

\begin{figure}
 \includegraphics[width=.5\textwidth]{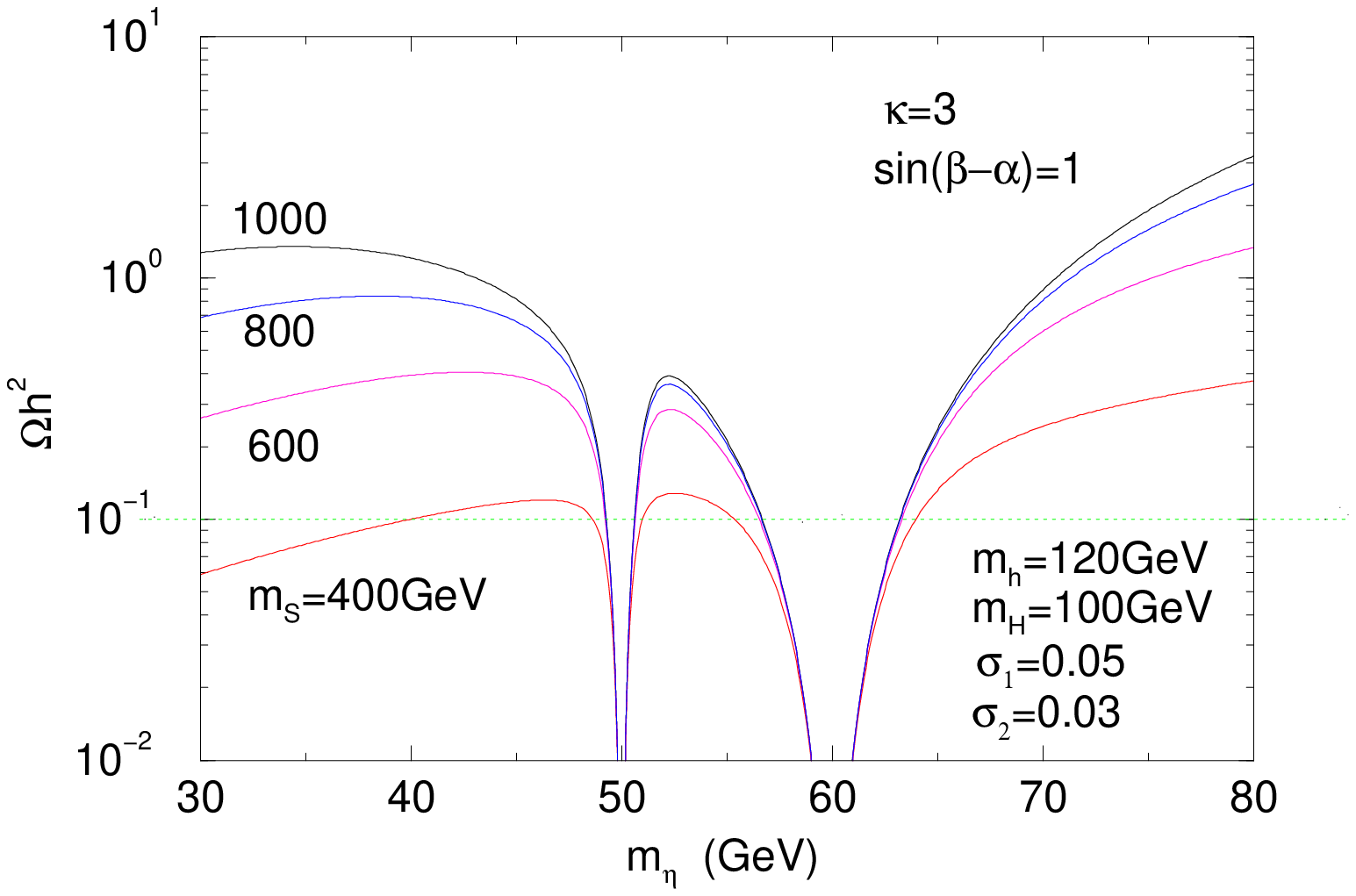}
 \includegraphics[width=.42\textwidth]{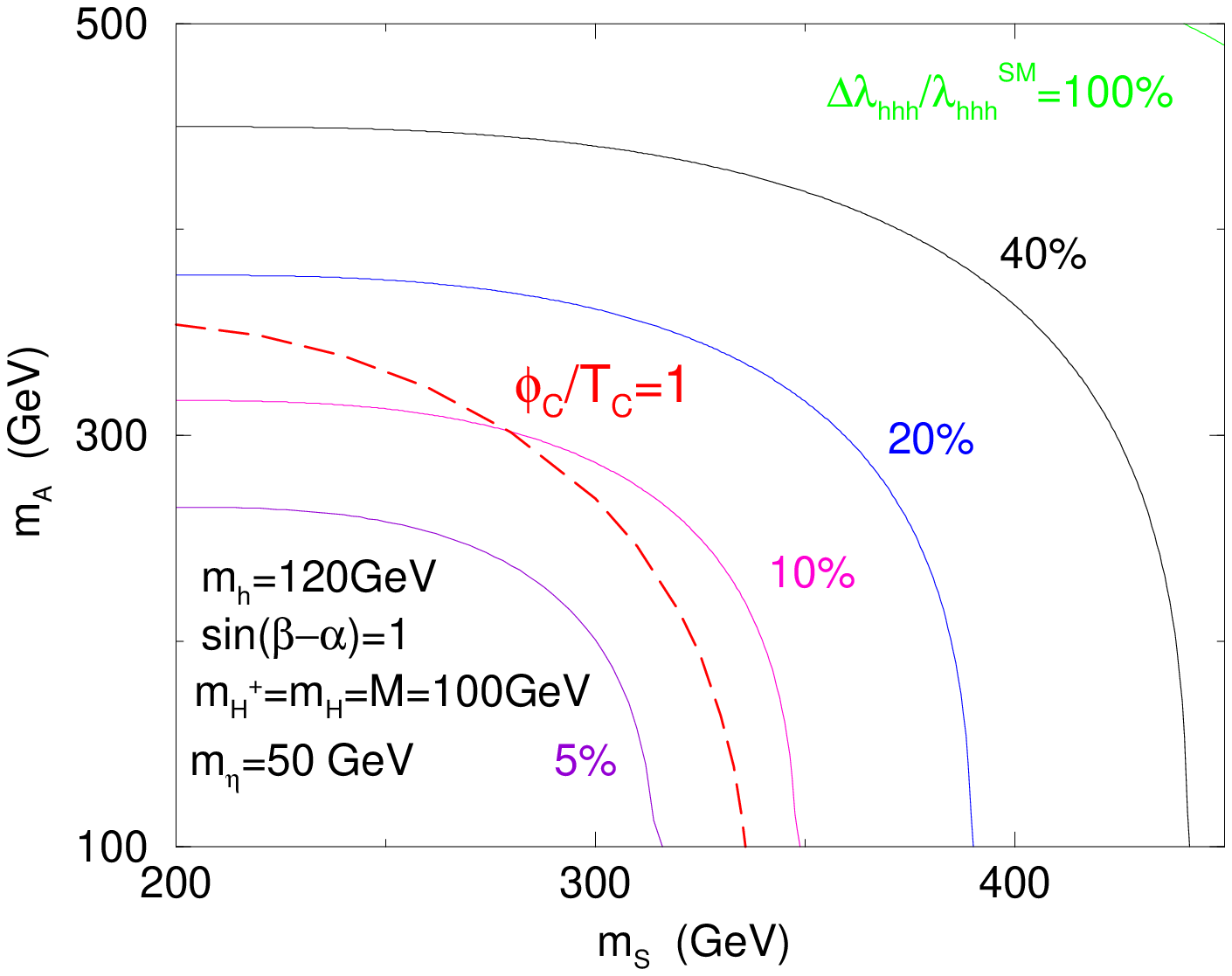}
  \caption{[Left figure] The relic abundance of $\eta$.
 [Right figure] The region of strong first order EWPT.
 Deviations from the SM value in the $hhh$ coupling
 are also shown. }
  \label{etaOmega}
\end{figure}

\indent
The lightest $Z_2$-odd particle is 
stable and can be a candidate of DM if it is neutral.
In our model, $N_R^\alpha$ must be heavy, so that 
the DM candidate is identified as $\eta$.
When $\eta$ is lighter than the W boson, $\eta$ dominantly annihilates 
into $b \bar{b}$ and $\tau^+\tau^-$ via tree-level $s$-channel
Higgs ($h$ and $H$) exchange diagrams, and into $\gamma\gamma$ via
one-loop diagrams.
From their summed thermal averaged annihilation rate $\langle \sigma v \rangle$,
the relic mass density  $\Omega_\eta h^2$ is 
evaluated.
Fig.~\ref{etaOmega}(Left) shows 
$\Omega_{\eta}h^2$ as a function of $m_\eta$. 
Strong annihilation can be seen near $50$ GeV $\simeq m_H^{}/2$
($60$ GeV $\simeq m_h/2$) due to the resonance of $H$ ($h$) mediation.
The data ($\Omega_{\rm DM} h^2 \simeq 0.11$~\cite{wimp}) indicate that $m_\eta$ is around 40-65 GeV. 
%


The model satisfies the necessary 
conditions for baryogenesis~\cite{sakharov}.
Especially, departure from thermal equilibrium can be 
realized by the strong first order EWPT.
The free energy is given at a high temperature $T$ by
\begin{eqnarray}
 V_{eff}[\varphi, T]= D (T^2-T_0^2) \varphi^2 
                     - E T \varphi^3 
                     + \frac{\lambda_T}{4} \varphi^4 + ..., 
\end{eqnarray}
where $\varphi$ is the order parameter.
A large value of the coefficient $E$
is crucial for the strong first order EWPT with keeping
$m_h \lsim 120$ GeV. 
For sufficient sphaleron decoupling in the broken phase, it is required that~\cite{sph-cond} 
\begin{eqnarray}
 \frac{\varphi_c}{T_{c}}  \left(\simeq \frac{2 E}{\lambda_{T_c}}\right) 
   \gsim 1, \label{sph2}
\end{eqnarray}
where $\varphi_c$ ($\neq 0$) and $T_c$ are the critical values of
$\varphi$ and $T$ at the EWPT.
In Fig.~\ref{etaOmega}(Right), the allowed region under the condition of
Eq.~(\ref{sph2}) is shown. The condition is satisfied
when
$m_{S^{\pm}}^{} \gsim 350$ GeV
for $m_A^{} \gsim 100$ GeV, 
$m_h \simeq 120$ GeV, $m_H^{} \simeq m_{H^\pm}^{} (\simeq 
M) \simeq 100$ GeV and $\sin(\beta-\alpha)\simeq 1$, where
$M$ represents the soft-breaking mass of extra Higgs bosons for
$\tilde{Z}_2$~\cite{aks}.

\section{Phenomenology}

A successful scenario which can simultaneously solve the above three issues 
under the data~\cite{lep-data,lfv-data,bsgamma} would be 
\begin{eqnarray}
 \begin{array}{llll}
 \sin(\beta-\alpha) \simeq 1, &\!\!
 \kappa \tan\beta \simeq 30, &\!\!
 m_h = 120 {\rm ~GeV},     &\!\!
 m_H^{} \simeq m_{H^\pm} \simeq {\cal O}(100) {\rm ~GeV},    \\
  m_A \gsim {\cal O}(100) {\rm ~GeV},
   &\!\! m_{S^\pm}^{}\sim 400{\rm ~GeV},&\!\!
  m_{\eta} \lsim m_W^{},
  &\!\! m_{N_R^{1}} \simeq m_{N_R^{2}} \simeq 3 {\rm ~TeV}.\\
  \end{array} \label{scenario}
\end{eqnarray}

\begin{wrapfigure}{r}{0.45\columnwidth}
\centerline{\includegraphics[width=0.37\columnwidth,angle=-90]{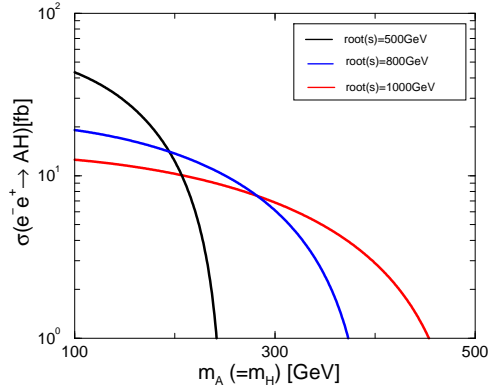}}
\caption{The production cross section of $e^+e^- \to HA$.}\label{Fig:eeHA}
\end{wrapfigure}

This is realized without assuming unnatural hierarchy among the
couplings. All the masses are  between ${\cal O}(100)$ GeV and ${\cal O}(1)$ TeV.
The discriminative properties of this scenario are in order: \\

\vspace{1.5mm}
      \noindent
      (I)~
      $h$ is the SM-like Higgs boson, but decays into $\eta\eta$ when $m_\eta < m_h/2$.
      The branching ratio is about 30\% for $m_\eta \simeq 43$ GeV and
      $\tan\beta=10$. 
      This is related to the DM abundance, so that our DM scenario is
      testable at the CERN Large Hadron Collider (LHC)
      and the ILC by searching the missing decay of $h$.
      Furthermore, $\eta$  is potentially detectable by direct DM searches~\cite{xmass},    
      because $\eta$ can scatter with nuclei via the scalar exchange~\cite{john}. 

\vspace{1.5mm}
      \noindent
      (II)~
      For successful baryogenesis, the $hhh$ coupling has to
      deviate from the SM value by more than 10-20
      \%~\cite{ewbg-thdm2} (see Fig.~\ref{etaOmega}), which can be tested
      at the ILC~\cite{hhh-measurement}.
    \begin{wrapfigure}{r}{0.45\columnwidth}
\centerline{\includegraphics[width=0.45\columnwidth]{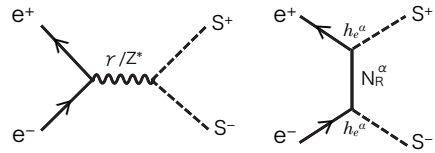}}
\caption{Feynman diagrams for the processes of  $e^+e^-\to
 S^+S^-$.}\label{fig:smsp_diag}
\end{wrapfigure}

\vspace{1.5mm}
      \noindent
      (III)~
      $H$ (or $A$) can predominantly decay into $\tau^+\tau^-$
      instead of $b\bar b$ for $\tan\beta\gsim 2$ because of the Type-X Yukawa interaction.     
      For example, we have 
      $B(H (A) \to\tau^+\tau^-) \simeq 100$ \% and $B(H (A) \to\mu^+\mu^-)  \simeq 
      0.3$ \% for $m_A^{}=m_H^{}=130$ GeV, $\sin(\beta-\alpha)=1$ and
      $\tan\beta=10$.
      The scenario with light $H^\pm$ and $H$ (or $A$) can be directly tested at the LHC
      via $pp\to W^\ast \to H H^\pm$ and $A H^\pm$~\cite{wah}, and also
       $pp \to HA$. Their signals are four lepton states
       $\ell^-\ell^+\tau^\pm\nu$ and $\ell^-\ell^+\tau^+\tau^-$, where
       $\ell$ represents $\mu$ and $\tau$~\cite{typeX}.
%
%
    At the ILC, the process $e^+e^- \to HA$ would be useful to discriminate
    the model from the other new physics candidates.  In 
    Fig.~\ref{Fig:eeHA}, the production rate of the $e^+e^- \to HA$
    is shown for $m_A^{} = m_H^{}$. 
    For $\sqrt{s}=500$ GeV, about 17,000 (110) of the $\tau^+\tau^-\tau^+\tau^-$
    ($\mu^+\mu^-\tau^+\tau^-$) events are then produced from the signal  
    for $m_A=m_H=130$ GeV~\cite{typeX},
    while about 60 (0) events are in the MSSM for the similar parameter set.
    The main back ground comes from $ZZ$ production (about 400 fb),
    which is expected to be easily reduced by appropriate kinematic cuts.

\vspace{1.5mm}
    \noindent
(IV)~
The physics of $Z_2$-odd charged singlet 
        $S^\pm$ is important to distinguish this model from the
        other models.
        At the LHC, they are produced in pair via the Drell-Yuan
        process~\cite{zee-ph}.
\begin{wrapfigure}{r}{0.45\columnwidth}
\centerline{\includegraphics[width=0.43\columnwidth]{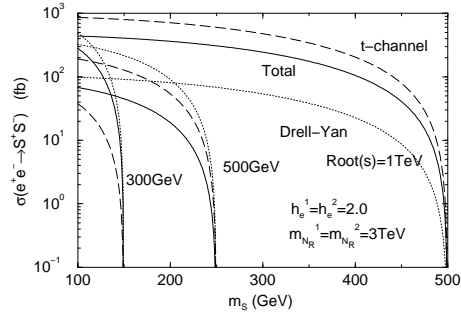}}
  \caption{ Production cross sections for $e^+e^- \to S^+S^-$ via the s-channel gauge boson
 ($\gamma$ and $Z$) mediation (dotted curve), the t-channel RH-neutrino
 ($N_R^\alpha$) mediation (solid curve), and both contributions (dashed
 curve) for $\sqrt{s}=300$, 500 and 1000 GeV. 
 } 
   \label{fig:smsp}
 \vspace{-3mm}
\end{wrapfigure}

        The cross section amounts to 0.5 fb for $m_{S^\pm}^{}=400$ GeV at $\sqrt{s}=14$ TeV, so that
        more than a hundred of the $S^+S^-$ events are produced for the
        integrated luminosity $300$ fb$^{-1}$. 
        The produced $S^\pm$ bosons decay as $S^\pm \to H^\pm \eta$, 
        and $H^\pm$ mainly decay into $\tau^\pm \nu$.
        The signal would be a high-energy hadron pair~\cite{hagiwara} with a large missing
        transverse momentum.

   The charged singlet scalar bosons $S^\pm$ in our model
   can also be better studied at the ILC via $e^+e^-\to S^+S^-$
   shown in Fig.~\ref{fig:smsp_diag}. 
   The total cross sections
    are shown as a function of $m_{S^\pm}^{}$ for
    $\sqrt{s}$
   in Fig.~\ref{fig:smsp}.
   The other relevant parameters are taken as 
   $m_{N_R^1}^{}=m_{N_R^2}^{}=3$ TeV and $h_e^1=h_e^2=2.0$.
   Both the contributions from the s-channel gauge boson ($\gamma$ and $Z$) mediation and the
   t-channel RH neutrino mediation are included in the calculation.
   The total cross section can amount to about 200 fb for
   $m_{S^\pm}^{}=400$ GeV at $\sqrt{s}=1$ TeV due to the contributions
   of the t-channel RH neutrino-mediation diagrams with ${\cal O}(1)$
   coupling constants $h_e^\alpha$.
   The signal would be a number of energetic tau lepton pairs with
   large missing energies. 
   Although several processes such as $e^+e^-\to W^+W^-$ and $e^+e^-\to H^+H^-$   
   can give backgrounds for this final state,
   we expect that the signal events can
   be separated by kinematic cuts.

\begin{wrapfigure}{r}{0.4\columnwidth}
\centerline{\includegraphics[width=0.3\columnwidth]{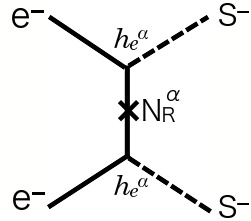}}
\caption{Feynman diagram for the processes of  $e^-e^-\to
 S^-S^-$.}\label{fig:smsm_diag}
 \vspace{-2mm}
\end{wrapfigure}

\begin{wrapfigure}{r}{0.45\columnwidth}
\centerline{\includegraphics[width=0.45\columnwidth]{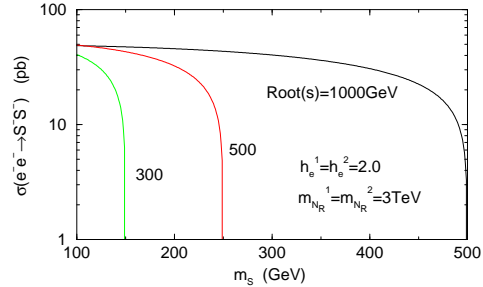}}
\caption{Production cross sections for $e^-e^- \to S^-S^-$ via the t-channel RH-neutrino
 ($N_R^\alpha$) mediation for $\sqrt{s}=300$, 500 and 1000 GeV.  
 }\label{fig:smsm}
 \vspace{-7mm}
\end{wrapfigure}

   Finally, there is a further advantage in testing our model at the
   $e^-e^-$ collision option of the ILC, where 
   the dimension five operators $\ell^- \ell^- S^+ S^+$, which appear in
   the sub-diagram of the three-loop induced masses
   of neutrinos in our model, can be directly measured.
   The production cross section for $e^- e^- \to S^-S^-$ [t-channel
   $N_R^\alpha$ mediation: see Fig.~\ref{fig:smsm_diag}] is given by
    \begin{eqnarray}
    \sigma(e^- e^- \to S^-S^-) = \int_{t_{\rm min}}^{t_{\rm max}}
     dt 
\frac{1}{128\pi s}  
      \left|\sum_{\alpha=1}^2
       (|h_e^\alpha|^2 m_{N_R^\alpha}^{})
       \left(\frac{1}{t-m_{N_R^\alpha}^2}+\frac{1}{u-m_{N_R^\alpha}^2}\right)\right|^2\, .
   \end{eqnarray}
     Due to the structure of our model that the tiny neutrino masses are
  generated at the three-loop level, the magnitudes of  $h_e^\alpha$
  ($\alpha=1,2$)
  are of ${\cal O}(1)$, by which the cross section becomes very large.
  Furthermore, thanks to the Majorana nature of the t-cahnnel diagram
  we obtain much larger cross section in the $e^-e^-$ collision than at
  the $e^+e^-$ collision when $m_{N_R^\alpha}^{2} \gg s$.
   Fig.~\ref{fig:smsm} shows the production cross sections for $e^-e^- \to S^-S^-$ via the t-channel 
  RH-neutrino.
The cross section can be as large as 30 pb for $m_{S^\pm}^{}=400$ GeV for
  $\sqrt{s}_{e^-e^-} = 1$ TeV, $m_{N_R^{1}}^{}=m_{N_R^{2}}^{}=3$ TeV and
  $h_e^1=h_e^2=2.0$. 
  The backgrounds are expected to be much less than the $e^+e^-$ collision. 

         We emphasize that a combined study for these processes
      would be an important test  for our model, in
  which neutrino masses are generated at the three-loop level by the 
  $Z_2$ symmetry and the TeV-scale RH neutrinos\footnote{Unlike our
      model, in the model in
      Ref.~\cite{knt}, the coupling constants corresponding to our
      $h_e^\alpha$ are small and instead those to $h_\mu^\alpha$ are
      ${\cal O}(1)$, so that its Majorana structure is not easy to test
      at $e^-e^-$ collisions.}. 
 
     In the other radiative seesaw models in which the neutrino masses are
  induced at the one-loop level with RH neutrinos, the corresponding
  coupling constants to our $h_e^\alpha$ couplings are necessarily
  one or two orders of magnitude smaller to satisfy the neutrino data, so that the cross
  section of the
  t-channel RH neutrino mediation processes are small due to the suppression 
  factor $(h_e^\alpha)^4$.

\vspace{1.5mm}
    \noindent
    (V)~
    The couplings $h_i^\alpha$ cause lepton flavor violation 
     such as $\mu\to e\gamma$ which would provide information
     on $m_{N_R^{\alpha}}$ at future experiments. 

\vspace{2mm}
Finally, we comment on the case with the CP violating phases.
Our model includes the THDM, so that the same discussion can be applied
in evaluation of  baryon number at the EWPT~\cite{ewbg-thdm}.
The mass spectrum  would be changed to some extent, but most of the features
discussed above should be conserved with a little modification. 

\section{Summary}

We have discussed the model with the extended Higgs sector and
TeV-scale RH neutrinos, which would explain neutrino mass and mixing,
DM and baryon asymmetry by the TeV scale physics. It gives
specific predictions on the collider phenomenology. In particular, 
the predictions on the Higgs physics are completely different from those in
the MSSM, so that the model can be distinguished at the LHC and also at
the ILC.

%

\begin{footnotesize}

\end{footnotesize}

\end{document}